# Iterative Algorithm for Finding Frequent Patterns in Transactional Databases

Gennady P. Berman[1], Vyacheslav N. Gorshkov[1], Edward P. MacKerrow[1], and Xidi Wang[2]

[1] Theoretical Division, T-13, MS-B213, Los Alamos National Laboratory, Los Alamos, NM, 87545
[2] xidi.wang@citicorp.com, Global Consumer Bank, Citigroup

A high-performance algorithm for searching for frequent patterns (FPs) in transactional databases is presented. The search for FPs is carried out by using an iterative sieve algorithm by computing the set of enclosed cycles. In each inner cycle of level $m$ FPs composed of $m$ elements are generated. The assigned number of enclosed cycles (the parameter of the problem) defines the maximum length of the desired FPs. The efficiency of the algorithm is produced by (i) the extremely simple logical searching scheme, (ii) the avoidance of recursive procedures, and (iii) the usage of only one-dimensional arrays of integers.

## 1. Introduction

Vast amount of consumer transactional details are being captured daily, which describe the trajectories of consumer behaviors, e.g. credit card transactions. It is well known observation that past behaviors predict future behaviors. Credit risk models are among the most widely used to rely on this observation to predict the future risk base on the past behaviors. In fact varies kind of models are commonly used for making daily decision throughout credit cycle in the card business, ranging from new customer acquisition, account maintenance, and collection queuing etc. The complexity of the models is necessary to be equal or more sophisticated than the corresponding behaviors they are predicting. Simple logistic regressions are often accurate enough for constructing several types of the models including the widely used credit risk models. The interactions among the commonly used variables only provide marginal lift over models without interactions. Such risk models are used in making decisions such as credit approval and credit line management, etc. and applied to the bulk part of the customer base.

There are situations that the behaviors of customers are much more sophisticated than a simple model can describe where important information is hidden among the subtle interactions/correlations among the variables. An example of which is case of the credit card fraud detection model, where the interactions among the transaction variables provide important clues about transactions being made, namely, fraudulent or not. By comparing the non-interactive models against the models build with considering all possible interactions one can find the model improvement due to the multivariate interactions. Credit card fraud detection rate benchmark indicates multifold improvement with the employment of interactive models over simple regressions. The multivariate interactions usually involve several variables, often expressed in form of expert rules, commonly used in the Market Basket Analysis (MBA) [1-6], e.g. *if under condition A and B and C and D, then the probability of Z occurs is X*, where the conditions A, B, C and D form a subspace (subsequently call a *pattern*) spanned by the input variables, and such subspace constitutes the interaction of involved variables. (As in the case of Citigroup's fraud detection rules developed, the number of variables involved is ranging from 4 to 12. The higher-order interactions are important in providing high-precision rules for the detection of frauds by minimizing false positives.) It is impractical for the statistical regression methods to extract such high-order interactions. It is also rather difficult to approximate such correlations through common parametric models such as neural networks.

The key strength of MBA lies in its capability of discovering the *best rules* under a given criteria (e.g. in the case of fraud detection, the best rule is defined as the rule with the highest probability of frauds occur at least 10 times in 3 months). The algorithm searches within the input space *exhaustively* to insure that there are no better rules to be missed. Such an exhaustive search algorithm can be very expensive computationally, since the number of the possible patterns grows exponentially with dimension of the input variables. By noticing that not all possible combinations of inputs occur for an actual database, and particularly if one is only interested in those patterns which appear at least above some minimum frequency threshold $\xi_0 \geq 1$, namely, *frequent patterns* (FPs) [5,6], then search space can be greatly reduced.

Mathematically this set of items can be represented as a transaction vector $\vec{T}$ in an *M*-dimensional space (where $M$ is the total number of items). The set of transactions $\vec{T}$ forms the database $B_0$. Each database includes many patterns, which are represented by the sets of items. Usually the number of these patterns is very large. The FP is defined in the following way. First, one chooses an arbitrary occurrence threshold $\xi_0 \geq 1$. If a given pattern appears in the database $f$ times, and $f \geq \xi_0$, then this pattern is a frequent pattern. There are two characteristic features of the FPs. One is the length of the FP. This length can vary from one up to the size of the transaction. Second, the maximum length of the FP depends on the threshold $\xi_0$, decreasing as $\xi_0$ increases. The FPs contain useful information about complicated correlations and rules hidden in the database. So, the extraction of FPs is an important task.

Because the search for FPs is a very important task, the creation of new fast approaches for finding FPs is essential. A number of methods [5, 7-10] have been devoted to the construction of "trees" for mining FPs. In recent work [6], a quite elegant alternative algorithm was presented which reduces the FPs search time and avoids the construction of these trees. However, in our opinion, the logical structure of the algorithm in [6] is overly complicated because the algorithm avoids the construction of the conditional databases in order to save the memory. Perhaps, this saving doesn't really take place because the initial database $B_0$ is represented by using non basic for C++ class of variables. Namely, every occurrence of item is stored in an entry with two fields: an item-*id* and hyper-*link* [6].

The algorithm proposed in our paper is implemented in Fortran-90 as well as in C++. It is characterized by (i) its simple logical structure, (ii) its use of only the integer variables in the programming languages, and (iii) its avoidance of recursive procedures. For generated in our simulations databases (see below), the operating speed of our algorithm is about 2-3 times faster than that of [6]. (We implemented the later in C++ in accordance with the recommendations given in [6].)

## 2. The general algorithm for searching for patterns

The transactions that form the database can be of two types. In the simplest case, each transaction is characterized by some "scalar" elements (variables) $X_i$ $(i=1,...,L)$ from the basic set. The number of elements, $k_i$, in the transaction, $\vec{T}_i$, is arbitrary $(k_i \leq L)$. One example of the transactional databases is the basket database used in the super-market analysis. Another type of transaction is the record type, for which each variable $X_i$ can take several independent values - $x_{i1}, x_{i2}, ..., x_{in_i}$. In this case, each transaction is a set of $M$ elements (one representative from each $X_i = \{x_{i1}, x_{i2}, ..., x_{in_i}\}$, $i = 1,......,M$). There is no fundamental difference in the FP search for either type of elements, because transactions of the second type can be easily reduced to the transactions of the first type. Namely, if we take $y_k$, $k = 1, 2, ..., K$, $K = \sum_{i=1}^{M} n_i$ as the basic elements ($y_1 \equiv x_{11}, y_2 \equiv x_{12}, ..., y_{n_1+1} \equiv x_{21}, ..., y_K \equiv x_{Mn_M}$), then both types of transactions can be represented in identical form.

Number of transaction
1 2 3 4 5 6 7 8 9 10 ……⇒………… 20

$x_1$: 0 0 1 1 1 0 1 0 1 1 0 1 1 1 1 0 0 1 1 1
$x_2$: 1 1 0 0 0 1 0 1 0 0 1 0 0 0 0 1 1 0 0 0
$x_3$: 0 0 1 0 1 0 0 0 1 1 0 1 1 1 0 0 0 1 1 1
$x_4$: 1 1 0 1 0 1 1 1 0 0 1 0 0 0 1 1 1 0 0 0
$x_5$: 0 1 0 0 0 1 0 1 0 1 0 1 1 0 0 1 0 0 0 1
$x_6$: 1 0 1 1 1 0 1 0 1 0 1 0 0 1 1 0 1 1 1 0
$x_7$: 0 0 **1 1 1 1 1 1 1** 0 **1 1 1 1** 0 0 **1 1 1 1**
$x_8$: 1 1 |0 0 0 0 0 0 0| 1 |0 0 0 0| 1 1 |0 0 0 0|
$x_9$: 0 0 |1 1 1 0 0 1 0| 0 |1 1 0 0| 0 1 |1 0 0 0|
$x_{10}$: 1 1 |0 0 0 1 1 0 1| 1 |0 0 1 1| 1 0 |0 1 1 1|
$x_{11}$: 0 1 |0 1 1 0 1 1 1| 0 |1 1 0 1| 0 0 |1 1 0 0|
$x_{12}$: 1 0 |1 0 0 1 0 0 0| 1 |0 0 1 0| 1 1 |0 0 1 1|
$x_{13}$: 1 1 |0 1 1 0 0 0 0| 0 |1 1 0 0| 1 0 |0 0 0 0|
$x_{14}$: 0 0 |1 0 0 1 1 1 1| 1 |0 0 1 1| 0 1 |1 1 1 1|

Table 1. Representation of the database $B_0$ by a matrix with $M$=14 basic elements. Every column represents a single transaction. The presence of 1 (0) in the row $m$ indicates that the element with the number $m$ is present (is not present) in this transaction. The three parts of the database $B_0$, contained in the rectangles, form the conditional database $B_7$ to the reference element $x_7$.

In what follows, we will consider databases $B_0$ in their general form (first type), constructed of basic elements $\{x_1, x_2, ..., x_M\}$. We represent this database as a matrix, with each column describing a transaction. (One (zero) in the column means the presence (absence) of the corresponding elements in the complete list of elements indicated in the left column in Table 1.)

Following traditional definitions [5], we define a database, $B_m$, which is conditional upon the element, $x_m$. (The element $x_m$ is called the "reference element".) The conditional database, $B_m$, is contained in the database $B_0$ and is constructed according to the following rules: Choose from the database, $B_0$, all columns that contain the element $x_m$ (in the row $m$). Then we remove all rows with numbers $i \leq m$.

The set of all possible patterns of elements for the arbitrary database, $B_k$, will be denoted by $[B_k]$. By a pattern we denote the tensor product $x_l x_j ... x_k$. The frequency of this pattern in the database, $B_k$, we define as $f_{lj...k}$. Then, the set of all patterns, $[B_k]$, can be represented as a sum of $f_{lj...k}(x_l x_j ... x_k)$.

We present the fundamental recurrent relation used for the construction of the algorithm proposed in this paper for searching for frequent patterns.

We will group all patterns using the following rule: first we isolate in $B_0$ all patterns which contain $x_1$; then (in the remaining set of patterns) we separate all patterns containing $x_2$, and so on, until we reach $x_M$ upon which we perform the final stage of the procedure. Then, $[B_0]$ can be represented as follows:

$$[B_0] = \sum_{k=1}^{M} x_k \left( f_k^{(B_0)} + [B_k] \right), \qquad (1)$$

where $f_k^{(B_0)}$ are the frequencies of $x_k$ in $B_0$.

Hence, the first level of operations in our algorithm consists of two parts: (i) the determination of the frequencies $f_k^{(B_0)}$ (frequencies of patterns of length 1) and (ii) the construction of the conditional databases $B_k$ (by performing a cycle on index $k$).

For the second level of operations in our algorithm we apply relation (1) to the databases $B_k$ (by performing a cycle on index $j$, which is included in the cycle of level 1):

$$[B_k] = \sum_{j=k+1}^{M} x_j \left( f_j^{(B_k)} + \left[ (B_k)_j \right] \right), \quad 1 \leq k \leq M-1. \qquad (2)$$

As a result, we will find all $f_j^{(B_k)}$, the frequencies of patterns, $x_i x_j$, $i < j$[*], of the length 2 and get all the databases $(B_k)_j$ for the operations on the third level, which will be enclosed in the level 2 cycle. By applying the relation (1) to the database for each level K, we obtain all patterns of length K, their frequencies, and the conditional databases for level K+1.

*The rules for extracting patterns and their frequencies:*

Every level $i$ of the system of enclosed cycles $(1 \leq i \leq L)$ generates patterns of length $i$. The list of elements contained in these patterns is the list of reference elements at the levels $1, 2, ..., i$. The frequencies of these patterns in the initial database are equal to the frequencies of the reference elements at level $i$ in the conditional database obtained at the level $(i-1)$. The maximum length of the patterns for L enclosed cycles is L.

*Frequent patterns*: If the number of the patterns in the database (the frequency, $f$) exceeds a given threshold, $\xi_0$ ($f \geq \xi_0$), then all these patterns are called *frequent patterns* (FPs). In order to find these FPs, we can exclude "infrequent" elements (with $f < \xi_0$) from all conditional databases. This procedure greatly reduces the calculation time. The greatest benefit of this procedure is achieved if the elements in all conditional databases are ordered from the lowest frequency to the highest: $f_i < f_m$, when $i < m$.

This ordering was used in the FP search algorithms based on tree construction [5]. This ordering was also used in our algorithm. Besides, we also build conditional databases, as it is done in [6]. If conditional databases are saved in the main memory in the form of integer arrays, the total size of conditional databases used in our algorithm does not exceed one needed for realization of the algorithm in [6]. We explain this observation by making a simple estimate. Suppose that we work with a database in which the probability of appearance of each element is 0.5. In this case, all conditional databases do not require more memory than the initial database $B_0$, if the database, $B_0$, is a one-dimensional array of integer numbers. If, for example, we consider transactions of the record type (for which each variable, $X_i$, can take several self-excluding values – $x_{i1}, x_{i2}, ..., x_{in_i}$, as defined above), then the full size of all additional databases relative to $B_0$ will become even smaller. The variables used in [6] require twice memory compared with the representation of $B_0$ by integers, because every item of the database $B_0$ is stored in an entry with two fields. (See above.) The efficiency of our algorithm is produced by (i) the extremely simple logical searching scheme, (ii) the avoidance of recursive procedures, and (iii) the usage of only one-dimensional arrays of integers.

---

[*] Such patterns are generated after substitution of (2) in (1)

Below we discuss the key points of our algorithm.

## 3. Basic technical methods for algorithm realization

| | |
|---|---|
| $x_1$: | 3, 4, 5, 7, 9, 10, 12, 13, 14, 15, 18, 19, 20 |
| $x_2$: | 1, 2, 6, 8, 11, 16, 17 |
| $x_3$: | 3, 5, 9, 10, 12, 13, 14, 18, 19, 20 |
| $x_4$: | 1, 2, 4, 6, 7, 8, 11, 15, 16, 17 |
| $x_5$: | 2, 6, 8, 10, 12, 13, 16, 20 |
| $x_6$: | 1, 3, 4, 5, 7, 9, 11, 14, 15, 17, 18, 19 |
| $x_7$: | 3, 4, 5, 6, 7, 8, 9, 11, 12, 13, 14, 17, 18, 19, 20 |
| $x_8$: | 1, 2, 10, 15, 16 |
| $x_9$: | 3, 4, 5, 8, 11, 12, 16, 17 |
| $x_{10}$: | 1, 2, 6, 7, 9, 10, 13, 14, 15, 18, 19, 20 |
| $x_{11}$: | 2, 4, 5, 7, 8, 9, 11, 12, 14, 17, 18 |
| $x_{12}$: | 1, 3, 6, 10, 13, 15, 16, 19, 20 |
| $x_{13}$: | 1, 2, 4, 5, 11, 12, 15 |
| $x_{14}$: | 3, 6, 7, 8, 9, 10, 13, 14, 16, 17, 18, 19, 20 |

Table 2. The representation of the database shown in Tab. 1 in the computer memory. For each element (the left column) there is a list of numbers of transactions in which this element is present.

The procedure of finding FPs reduces to the construction of conditional databases on the level (*m+1*) for the reference elements from the database constructed on the level *m*. The frequencies of elements in these conditional databases are determined simultaneously with the construction of these databases. As an illustration, we consider the database in Table 1 as the level *m* database.

The database in the computer memory is introduced as lists of numbers of transactions that contain a given basic element. (See Table 2.) These lists form the one-dimensional array of integers, $B_m(l)$. The length of this array is equal to the number of non-zero elements in all the transactions of the database in Table 1. There are three key-arrays necessary to work with the database shown in Table 2. These keys are shown in Table 3. Note that the database $B_m(l)$ itself is not ordered according to the increasing frequency of the elements. (This is important to minimize the FP search time.) All ordering functions are imposed on the $Rate_m(i)$-array in Table 3. When the "formal parameter" $i$ of the cycle of level $m$ increases, we work with the elements which are determined in compliance with the "real parameter" $j \equiv Rate_m(i)$. This real parameter, $j$, is used to select elements in the total list of elements which is indicated as $Name_m(j)$ in Table 3. The parameter $j$ also indicates the necessary starting addresses of the list of transactions $Adr_m(j)$ in $B_m(l)$) in the order of increasing frequencies, $f_m(j)$.

| The list of the database elements' names on level $m$: $Name_m(j)$ | The frequencies of the elements in the database $f_m(j)$ | Initial addresses of the list of transactions for each element in $B_m(l)$: $Adr_m(j)$ | The pointers to the numbers of elements in the list of names in order of increasing frequency: $Rate_m(i)$ | The values of the formal parameter on level $m$: $i$ |
|---|---|---|---|---|
| \multicolumn{5}{c}{$j = Rate_m(i)$} | | | | |
| $x_1$ | 13 | 1 | 8 | 1 |
| $x_2$ | 7 | 14 | 13 | 2 |
| $x_3$ | 10 | 21 | 2 | 3 |
| $x_4$ | 10 | 31 | 9 | 4 |
| $x_5$ | 8 | 41 | 5 | 5 |
| $x_6$ | 12 | 49 | 12 | 6 |
| $x_7$ | 15 | 61 | 4 | 7 |
| $x_8$ | 5 | 76 | 3 | 8 |
| $x_9$ | 8 | 81 | 11 | 9 |
| $x_{10}$ | 12 | 89 | 10 | 10 |
| $x_{11}$ | 11 | 101 | 6 | 11 |
| $x_{12}$ | 9 | 112 | 14 | 12 |
| $x_{13}$ | 7 | 121 | 1 | 13 |
| $x_{14}$ | 13 | 128 | 7 | 14 |

Table 3. Key-arrays for the database in Tab. 2. $f_m(j)$, $Adr_m(j)$, $Rate_m(i)$ are the one-dimensional arrays of integers. Their lengths are equal to the number of elements on each level $m$. $Name_m(j)$ is an array of character type, but can be an integer, if the elements' names are coded by integers in the initial database $B_0$. The sequence order of elements is not important if they have the same frequency.

Suppose that the value of the current formal parameter $i$ on the level $m$ is $i = 6$. Then, in accordance with $Rate_m(6) = 12$, the reference element for the conditional database, which is transmitted to the level ($m+1$), is $x_{12}$ with the frequency 9*. This element is present in the following transactions 1, 3, 6, 10, 13, 15, 16, 19, 20. (See Table 1.) The elements that occur in the conditional database can be obtained from the values, $Rate_m (7 \leq i \leq 14)$: $j = 4, 3, 11, 10, 6, 14, 1, 7$. A selection of the names from the $Name_m(j)$-array (using these values of $j$), gives us the set of 8 elements - $\{x_4, x_3, x_{11}, x_{10}, x_6, x_{14}, x_1, x_7\}$ in the conditional database transformed to the level ($m+1$). Now, in the lists of transactions for these elements we have to take only those numbers of transactions that occur in the list of transactions for the element $x_{12}$.

For example, if the threshold is $\xi_0 = 5$, we will obtained the following conditional database. (See Tab. 4.)

The key-arrays for this database are presented in Table 5. We present the details of the construction of the database $B_{m+1}(l)$ and the corresponding arrays in Table 5.

To compare the list of transactions of the reference element in the database $B_m(l)$ with the lists of transactions of other elements, we perform the following procedure. We create the one-dimensional auxiliary array of integers called

| $x_3$: | 3 ,10,13,19,20 |
| $x_{10}$: | 1 , 6 ,10,13,15,19,20 |
| $x_{14}$: | 3, 6, 10,13,16,19,20 |
| $x_1$: | 3,10,13,15,19,20 |
| $x_7$: | 3 , 6, 13,19,20 |

Table 4. The $B_{m+1}(l)$ database of the level ($m+1$) for the current reference element $x_{12}$ at the level $m$. (The formal parameter of the cycle on the level $m$ is $i = 6$.)

| The list of the database elements' names on level $m$: $Name_{m+1}(j)$. | The frequencies of the elements in the database $f_{m+1}(j)$. | Initial addresses of the list of transactions for each element in $B_m(l)$: $Adr_m(j)$ | The pointers to the numbers of elements in the list of names in order of increasing frequency: $Rate_m(i)$ | The values of the formal parameter on level $m$: $i$ |
|---|---|---|---|---|
| $x_3$ | 5 | 1 | 1 | 1 |
| $x_{10}$ | 7 | 6 | 5 | 2 |
| $x_{14}$ | 7 | 13 | 4 | 3 |
| $x_1$ | 6 | 20 | 2 | 4 |
| $x_7$ | 5 | 26 | 3 | 5 |

Table 5. Key-arrays for the database from Tab. 4.

$KodTr(k)$. The length of this array is equal to the number of transactions in $B_0$. This array is common for all cycles on all levels. Initially it contains all zeros. Before constructing the database, $B_{m+1}(l)$, we scan the list of transactions for the reference element in the database $B_m(l)$: $Adr_{1sup} \leq l \leq Adr_{2sup}$, $Adr_{1sup} = Adr_m(j_{sup})$, $Adr_{2sup} = Adr_m(j_{sup}) + f_m(j_{sup}) - 1$ (where $j_{\sup} \equiv Rate_m(i)$ is the real parameter of the reference element; $i$ is the current value of the formal parameter at the level $m$). The numbers of transactions $NumTr = B_m(l)$, reading sequentially, are used as indexes of the array $KodTr(k)$ for ascribing to these elements the value one: $KodTr(NumTr) = 1$. The minimal and maximum values of $NumTr$ are: $NumTr_{\min}^{(\sup)} = B_m(Adr_{1\sup})$, $NumTr_{\max}^{(\sup)} = B_m(Adr_{2\sup})$.

---

* We recall that for the current formal parameter $i = 6$ on level $m$, the FP of length $m$ will be generated. Its frequency is 9, and a list of elements consists of the list of reference elements on all levels from 1 to $m$. The last element is $x_{12}$.

In our case, when the value of the formal parameter $i$ at level $m$ is $i = 6$, for the reference element $x_{12}$ the $KodTr(k)$ array becomes:

| $k$ | **1** | 2 | **3** | 4 | 5 | **6** | 7 | 8 | 9 | **10** | 11 | 12 | **13** | 14 | **15** | **16** | 17 | 18 | **19** | **20** |
|---|---|---|---|---|---|---|---|---|---|---|---|---|---|---|---|---|---|---|---|---|
| $KodTr(k)$ | **1** | 0 | **1** | 0 | 0 | **1** | 0 | 0 | 0 | **1** | 0 | 0 | **1** | 0 | **1** | **1** | 0 | 0 | **1** | **1** |
| $NewTr(k)$ | **1** | - | **2** | - | - | **3** | - | - | - | **4** | - | - | **5** | - | **6** | **7** | - | - | **8** | **9** |

(The integer array $NewTr(k)$ is jointly built with the array $KodTr(k)$, and it represents the *new numbers* of transactions in the conditional database.)

The objective of this procedure is to prepare the template of the list of transactions for the reference element, reading this list only once. The list of transactions for other elements will be compared with this array.

Now, we describe the construction of the conditional database. We take the elements with the formal parameter $it$: $14 \geq it > i = 6$. The first such element, according to the actual parameter $jt = Rate_m(it = 7) = 4$, is the element that holds the fourth position in the list of names $Name_m(jt)$. In this case, its name is $x_4$, and all its characteristics will occur in the conditional database $B_{m+1}(l)$ in the first positions in the key-arrays. We will fill these arrays using the fill-index $Inf$ with the starting value $Inf = 1$. The finite value of this index is equal to the number of elements in the constructed conditional database $B_{m+1}(l)$. So, the first steps are: $Name_{m+1}(Inf) = x_4$, $Adr_{m+1}(Inf) = 1$. After this we read the numbers of transactions in the list for $x_4$: $NumTr = B_m(l)$, $Adr_m(jt) \leq l \leq Adr_m(jt) + f_m(jt) - 1$ ($jt=4$). We check to see if there is such a number in the list of transactions for the reference element[*], using the auxiliary array $KodTr$. If $KodTr(NumTr) = 1$ (there is a coincidence), then we include this transaction number $NumTr$ in the list of transactions for the element $x_4$ in the conditional database (starting from the first element of array $B_{m+1}(l)$, because $Adr_{m+1}(Inf) = 1$). At the same time, we increase the counter for the frequency, $f$, of element $x_4$ by one, and so on. After finishing all operations with the element $x_4$, we have to check that the condition, $f \geq \xi_0$, is fulfilled. If so, we should store the value of $f$ in $f_{m+1}(Inf)$, define $Adr_{m+1}(Inf + 1) = Adr_{m+1}(Inf) + f_{m+1}(Inf)$, and increase the fill-index by 1: $Inf = Inf + 1$. After this, we move to the analysis of the next element $x_3 (it = 8)$, whose characteristics will be placed into the second positions in the key-arrays for $B_{m+1}(l)$. However, in our example, the frequency of $x_4$ is less than the threshold $\xi_0$. In this and similar cases, we move on to the analysis of the next element without changing the fill-index $Inf$, excluding the previous element from the database $B_{m+1}(l)$.

---

[*] It is natural that we do not perform this procedure for $NumTr < NumTr_{\min}^{(\sup)}$, and when $NumTr > NumTr_{\max}^{(\sup)}$ we stop the search for the same numbers in the transactions.

As a result, we obtain the final representation of $B_{m+1}(l)$ (see Table 4) and its key-arrays $Name_{m+1}(j)$, $f_{m+1}(j)$, $Adr_{m+1}(j)$, $j = 1, 2, ..., 5$. The final stage is the construction of $Rate_{m+1}(i)$. For this, the frequencies $f_{m+1}(1), f_{m+1}(2), f_{m+1}(3), f_{m+1}(4), f_{m+1}(5)$ have to be ordered in their increasing values. Finally, we will obtain the following sequence: $f_{m+1}(1), f_{m+1}(5), f_{m+1}(4), f_{m+1}(2), f_{m+1}(3)$. The indexes of this ordered array represent the values of $Rate_{m+1}(i)$: $Rate_{m+1}(i) = 1, 5, 4, 2, 3$ for $i = 1, 2, 3, 4, 5$. In accordance with this order, the temporary cycle at level $(m+1)$ will produce FPs of length $(m+1)$: $5......x_{12}x_3$; $5......x_{12}x_7$; $6......x_{12}x_1$; $7......x_{12}x_{10}$, $7......x_{12}x_{14}$.

To decrease the total size of conditional databases it is desirable to change a numeration of transactions in those databases in accordance with the integer array $NewTr(k)$. These changes are performed in the process of construction of the conditional database. Also, the values $\hat{B}_m(l) \equiv B_{m+1}(l) - B_{m+1}(l-1)$ have to be stored in the list of transactions for each item $x_j$ of the database $B_{m+1}$. Note, that the following condition on index $l$ must be satisfied: $Adr_{m+1}(j) + f_{m+1}(j) \geq l \geq Adr_{m+1}(j) + 1$. The initial numbers in the list of transactions remains unchanged: $\hat{B}_m(Adr_{m+1}(j)) \equiv B_{m+1}(Adr_{m+1}(j))$. For example, the conditional database $B_{m+1}$ (Table 4) could have been represented by means of the Table 6a, if the new numeration of transactions were used in accordance with the array $NewTr(k)$. But, to decrease the total size of this databases, we really store in the main memory the values $B_{m+1}(l) - B_{m+1}(l-1)$ instead of $B_{m+1}(l)$, so this database $B_{m+1}$ can be presented by Table 6b.

| $x_3$: 2, 4, 5, 8, 9 |
|---|
| $x_{10}$: 1, 3, 4, 5, 6, 8, 9 |
| $x_{14}$: 2, 3, 4, 5, 7, 8, 9 |
| $x_1$: 2, 4, 5, 6, 8, 9 |
| $x_7$: 2, 3, 5, 8, 9 |

Table 6a. Representation of conditional database $B_{m+1}$ by using the new numeration of transactions.

| $x_3$: 2, 2, 1, 3, 1 |
|---|
| $x_{10}$: 1, 2, 1, 1, 1, 2, 1 |
| $x_{14}$: 2, 1, 1, 1, 2, 1, 1 |
| $x_1$: 2, 2, 1, 1, 2, 1 |
| $x_7$: 2, 1, 2, 3, 1 |

Table 6b. Real representation of conditional database $B_{m+1}$ in main memory

This representation allows one to keep each item of any database using a minimum number of bytes in the main memory.

## 4. Non-Reducible FPs

As it was mentioned in the Introduction, the total number of FPs can be very large. At the same time, usually not all FPs are equally important for characterization of the database properties. We discuss here one of the possible approaches which can be used to extract the "non-reducible" FPs which characterize the database in a most representative way. Let's consider a simple example. Suppose that the FP $ab$ overcomes the threshold $\xi_0$ in the database $B_0$. However, the items $a$ and $b$ can be statistically independent. In this case, the FP $ab$ cannot be considered as the new characteristic (rule) of the database. So, we place at level 2 of our algorithm the "filter" for such reducible FPs.

The corresponding condition is

$$\left| \frac{p(ab)}{p(a)p(b)} - 1 \right| \leq \varepsilon . \tag{3}$$

Then, the FP $ab$ is excluded from further consideration (where $\varepsilon$ is some small number). Also, we will not search for other longer FPs which contain the combination $ab$ at deeper levels of the algorithm. In (3) $p(ab)$ is the probability of the FP $ab$ occurrence in the database $B_0$; $p(a)$ and $p(b)$ are the probabilities of FPs $a$ and $b$ of length 1; $p(ab) \approx f_{ab}/N$, $p(a) \approx f_a/N$, $p(b) \approx f_b/N$, where $N$ is the total number of transactions in the database, $f_{ab}$, $f_a$, $f_b$ are the frequencies of the corresponding FPs which are known at level 2.

Now we will estimate the value of $\varepsilon$, assuming that we have a set of similar databases, and the averaging can be performed over this set. If elements $a$ and $b$ are statistically independent, then the probability of the FP $ab$ is equal to $\hat{p}(ab) = p(a)p(b) \equiv q$. The corresponding frequency $\hat{f}_{ab}$ can be considered as a random variable with the average value $\overline{\hat{f}}_{ab} = Nq$. The dispersion of $\hat{f}_{ab}$ is equal to $D = N(q - q^2)$, and its root mean square is $\sigma = \sqrt{D}$. Then, the criteria of the statistical independence of random variables $a, b$ (3) could be presented as $\left| f_{ab} - \overline{\hat{f}}_{ab} \right| \leq 3\sigma$, or

$$\left| \frac{f_{ab}}{Nq} - 1 \right| \leq \frac{3\sqrt{1-q}}{\sqrt{Nq}} \equiv \varepsilon . \tag{4}$$

Filters like (4) can be easily placed at each level of the algorithm. If the FP $(a,b,...,z,e)$ of the length $m$ and frequency $f_{ab...ze}$ is obtained at level $m$, then we have to turn on the filter $\left| \frac{f_{ab...ze}}{Nq} - 1 \right| \leq \varepsilon$ with the parameter $\frac{f_{ab...z}}{N} \cdot \frac{f_e}{N}$, where $f_{ab...z}$ is the frequency of the FP $(a,b,...,z)$ at level $(m-1)$. This frequency is known when we proceed to level $m$. Thus, the structure of the proposed algorithm allows one to perform the filtration of non-reducible FPs. This procedure reduces significantly the time of FPs search.

## 5. Some final remarks

1. It is much more convenient to represent the elements of the database by integers for the following reason: suppose that we have a database of the second type, in which each variable $X_i$ can take up to 50 independent values $x_{i,k}$, $k = 1, 2, ...., 50$. We represent the elements' names as $Name(x_{i,k}) = 100 \cdot i + k$. For instance, $x_{25,3}$ will be coded as 2503. This is convenient while constructing the conditional database, because it is easy to determine the independent elements using the following characteristic: $|Name(x_{i1, k1}) - Name(x_{i2, k2})| < 50$, and to avoid the comparison of the list of transactions for $i1=i2$.

2. If the frequency of the element (or some $n_m$ elements) in the conditional database, transmitted to the level $(m+1)$, coincides with the frequency of the reference element on level $m$, we may exclude these elements from the conditional database. At the same time, we must form on the level $(m+1)$ FPs of length $(m+1+n_m)$ taking into account these excluded elements. For example, let $x_9$ be the reference element at the level $m$ with the frequency $f_m(x_9)$. In the conditional database, the elements $x_{34}, x_{21}, x_{64}$ have the same frequency. We exclude them from the conditional database, but we set these elements on the level $(m+1)$ as "additional" elements, Fig. 1. So, the level $(m+1)$ produces FPs of length $(m+3)$: $f_m(x_9).......x_9(x_{34}x_{21}x_{64})$ - the green arrow in Fig. 1. After this, we generate the FPs of length $(m+4)$, in accordance with the frequency and name of the current reference element in the cycle of the level $(m+1)$ - blue arrow in Fig. 1. For instance, if $x_5$ is the reference element at level $(m+1)$ with frequency $f_{m+1}(x_5)$, then the corresponding FP is $f_{m+1}(x_5).......x_9(x_{34}x_{21}x_{64})x_5$. Using this rule, we ignore FPs of length $(m+1)$ and $(m+2)$ like $f_m(x_9).......x_9x_{34}$, $f_{m+1}(x_5).......x_9x_5$ or $f_m(x_9).......x_9x_{34}x_{21}$, $f_{m+1}(x_5).......x_9x_{21}x_5$. However, they can be easily derived from the longer FPs, based on the typical group of elements, $x_9(x_{34}x_{21}x_{64})$. FPs which include the groups $x_9(x_{34}x_{21}x_{64})$ (usually they are long FPs) should be taken into consideration in the first place for classification of the initial database.

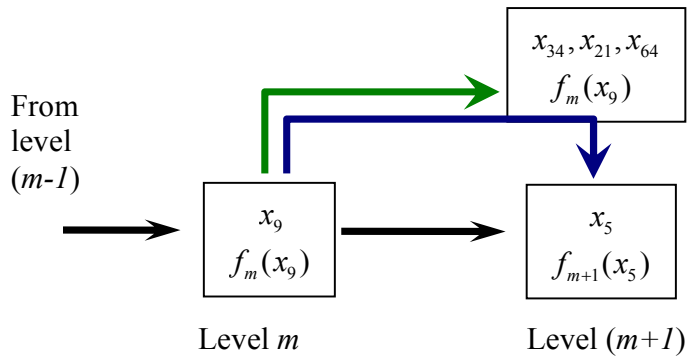

Fig. 1

3. For comparison of the effectiveness of our algorithm with other algorithms, we present the following test results. The search for FPs in a database with 100,000 transactions and 50 basic elements with the threshold, $\xi_0 = 4000$, takes 150 seconds (independent of the programming language: Fortran-90 or C++). When generating the transactions for the initial database, the probability of occurrence of each of the 50 basic elements in any transaction was 0.5. We used a 3.2 GHz DELL computer.

   4. *Dependence of the computation time on the length of the FPs*: Setting the number of levels *K* (included cycles), we define the maximum length of desired FPs. There is a relation between the computation time and the parameter *K* (Fig. 2) for a database with 10,000 transactions, 50 basic elements and the threshold, $\xi_0 = 20$. (The rules for the database construction were described in remark 3.) With the increasing K, the number of possible patterns whose frequencies have to be found, increases approximately as $\sum_{k=1}^{K} C_M^k$ (where M – number of basic elements). This is the major factor in determining the calculation time, $\sim \exp(K)$, if K is relatively small. At the same time, the number of transactions in the conditional databases decreases steadily as we move on to the deeper levels. If K is sufficiently large, the frequencies of the long FPs approach a threshold. In this case, the processing of the conditional databases takes minimal time, and the total processing time is practically saturated. These regularities have to be observed when processing databases of any kind.

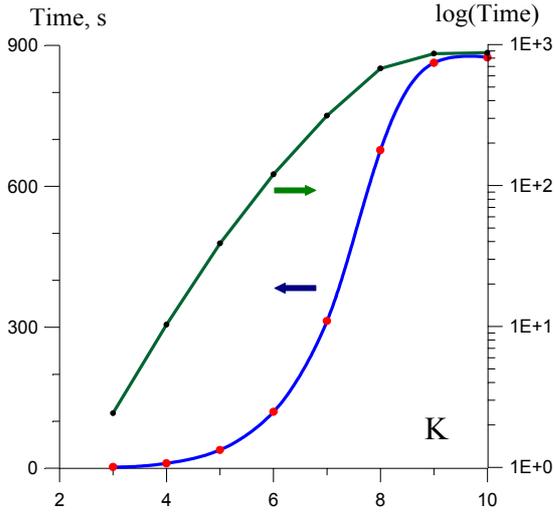

Fig. 2. The dependence of the computation time on the maximum length of the FPs (Blue curve); (green- in the log(Time)-scale).

## 6. Conclusion

The algorithm described in this paper has a high speed performance, as:

- In the process of algorithm execution neither ordering nor re-arrangement takes place directly in the database. Ordering is performed only for frequencies of the elements while constructing the key-arrays, $Rate_m(i)$.

- The construction of conditional databases for level $(m+1)$ is performed directly from the conditional databases obtained at level $m$. This avoids the time losses for constructing the various types of trees.

- There are no recursive operations (these slow down the computation process) and no non-basic variables for programming languages. The whole algorithm operates on one-dimensional arrays of integers.

- Filtering criteria by confidence and importance can be applied at each level of iteration such that useful patterns can be identified.

Note that the suggested algorithm is in many respects similar to the algorithm [6]. Distinctions are related to some technical realizations of computational process, which provide a reduction of the FPs search time.

## Acknowledgments

This work was supported by the Department of Energy under the contract W-7405-ENG-36 and DOE Office of Basic Energy Sciences. We also thank the ISR-DO at LANL for supporting this research.

## References.

[1] Agarwal, R., Agarwal, C., and Prasad, V.V.V. A tree projection algorithm for generation of frequent item sets. Journal of Parallel and Distributed Computing, **61**, no.3, 350-371 (2001).

[2] Jasper, E. Global query processing in the AutoMed heterogeneous database environment. Advances in Databases. 19th British National Conference on Databases, BNCOD 19. Proceedings, Sheffield, UK, 17-19 July 2002, p. 46-49.

[3] David Taniar, J. Wenny Rahayu. Parallel database sorting. Information Sciences, **146**, no. 1-4, 171-219 (2002).

[4] Savasere, A., Omiecinski, E., and Navathe, S. An efficient algorithm for mining association rules in large databases. Proc. 21st International Conference on Very Large Databases (VLDB), Zurich, Switzerland, 1995; Also Gatech Technical Report No. GIT-CC-95-04.

[5] Han, J.W, Pei, J., and Yin, Y.W. Mining frequent patterns without candidate generation: a frequent-pattern tree approach. Data Mining and Knowledge Discovery, **8**, no. 1, 53-87 (2004).

[6] Pei, J., Han, J.W., Lu, H.I., Nishio, S., Tang, S.W., Yang, D.Q. H-Mine: Hyper-structure mining of frequent patterns in large databases. IEEE International Conference on Data Mining, San Jose, CA, USA, 2001, p. 441-448.

[7] Xu, Y., Yu, J.X., Liu, G., Lu, H. From path tree to frequent patterns: a framework for mining frequent patterns. IEEE International Conference on Data Mining, 2002, p. 514-521.

[8] Cheung, W., Zaïane, O.R. Incremental mining of frequent patterns without candidate generation or support constraint. International Database Engineering and Applications Symposium, Hong Kong, China, 16-18 July 2003, p. 111-116.

[9] Fan, H.J., Fan, M., Wang, B.Z., Maximum item first pattern growth for mining frequent patterns. Lecture notes in artificial intelligence; **2639**, 515-523 (2003).

[10] Liu, G.M., Lu, H.J., Xu, Y.B., Yu, J.X. Ascending frequency ordered prefix-tree: efficient mining of frequent patterns. Eighth International Conference on Database Systems for Advanced Applications (DASFAA), Kyoto, Japan, 26-28 March, 2003, p. 65-72.

[11] Liu, G., Pan, Y., Wang, K., Han, J. Mining frequent itemsets by opportunistic projection. Proc. of CIKM Conference, 2000, p. 5-11.